\documentclass[a4paper, oneside, twocolumn, notitlepage, 10pt]{extarticle_ecoc}
\usepackage{ecoc}
\usepackage{booktabs}

\DeclareMathOperator*{\minimize}{minimize}
\DeclareMathOperator*{\argmin}{arg\,min}
\newcommand{\expop}{\mathbb{E}}
\newcommand{\testleft}{\leftarrow\!\shortmid}
\begin{document}
\selectlanguage{english}    

\title{Nonlinear Equalization for Optical Communications\\Based on Entropy-Regularized Mean Square Error}%


\author{
    Francesca Diedolo\textsuperscript{(1)}, Georg B\"{o}cherer\textsuperscript{(2)},
    Maximilian Sch\"adler\textsuperscript{(2)}, Stefano Calabr\'o\textsuperscript{(2)}
}

\maketitle                  


\begin{strip}
 \begin{author_descr}

   \textsuperscript{(1)} Technical University of Munich, Institute for Communications Engineering, Theresienstr. 90, 80333 Munich, Germany,
   \textcolor{blue}{\uline{francesca.diedolo@tum.de}} 

   \textsuperscript{(2)} Huawei Munich Research Center, Riesstr. 25, 80992 Munich, Germany

 \end{author_descr}
\end{strip}

\setstretch{1.1}
\renewcommand\footnotemark{}
\renewcommand\footnoterule{}
\newcommand{\tcr}[1]{\textcolor{red}{#1}}


\begin{strip}
  \begin{ecoc_abstract}
   An entropy-regularized mean square error (MSE-X) cost function is proposed for nonlinear equalization of short-reach optical channels. For a coherent optical transmission experiment, MSE-X achieves the same bit error rate as the standard MSE cost function and a significantly higher achievable information rate. 
   \textcopyright2022 The Author(s)
  \end{ecoc_abstract}
\end{strip}


\section{Introduction}
Transmission over optical fiber is characterised by nonlinear impairments such as the Kerr effect~\cite{Kikuchi16}. Additional nonlinearities are caused by imperfect transceiver (optical and electrical) devices, and  compensating these nonlinearities is crucial for short reach applications~\cite{Bluemm2019}.
Several nonlinearity compensation techniques have been studied~\cite{amari2017}, e.g., Volterra equalizers and equalizers based on neural networks (NNs)~\cite{Zibar16}. 
In~\cite{Jarajreh15, Zhang19, Freire21}, nonlinear equalizers have been optimized through a mean squared error (MSE) cost function, which is equivalent to the minimum MSE (MMSE) criterion, traditionally used for linear equalizers. This approach is optimal for hard-decision (HD) systems when there is additive noise with circularly-symmetric complex Gaussian statistics.

Modern optical communication systems rely on soft-decision (SD) forward error correction (FEC)~\cite{Alvarado15}. For such systems, the HD performance gives less insight than the achievable information rate (AIR) or the generalized mutual information (GMI)~\cite{Alvarado15}. 
The authors of~\cite{Deligiannidis20, Schadler21} noted that minimizing the MSE results in a grid-shaped scatterplot (MSE grid), see Fig.~\ref{fig:scatterplots}(b), that gives a poor AIR. To improve the AIR, nonlinear equalizers should instead be optimized according to average cross entropy (CE). The average CE is based on a demapper output and~\cite{Deligiannidis20, Schadler21} propose NNs that perform equalization and demapping jointly. A disadvantage of a joint approach is that one loses access to the equalized signal before the demapper.
This is important because some algorithms, e.g. carrier recovery and timing recovery, need access to the equalized signal.

In this work, we suggest entropy-regularized MSE (MSE-X) as a cost function. If the equalizer is followed by a demapper, which we define formally below, then MSE-X achieves the same BER as using the MSE cost function but a larger AIR. Moreover, MSE-X lets us separate equalization and demapping. Fig.~\ref{fig:scatterplots}(c) visualizes this through the scatterplot of MSE-X equalized signals.

This paper is organized as follows.
We first review cost functions and highlight their drawbacks for nonlinear equalization. We then derive the MSE-X cost function and evaluate it for an optical experiment to illustrate the superiority of MSE-X over MSE. 

\section{Design Criterium for SD-FEC}
\begin{figure}
    \centering
    \includegraphics[width=0.5\textwidth]{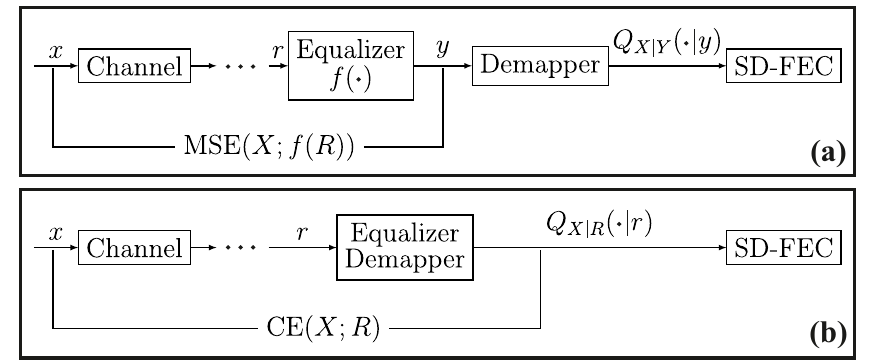}
    \caption{Transceiver model for SD-FEC  based communication systems with MSE training (a) and CE training (b).}
    \label{fig:transceiver_model}
\end{figure}
Consider the model depicted in Fig.~\ref{fig:transceiver_model}. A demapper provides a SD to the FEC decoder in the form of an a posteriori distribution $Q_{X|Y}(x|y)$, where $x\in\mathcal{X}$ and where $y$ is the demapper input. 

\textbf{Remark} Most systems use binary FEC and the demapper output is a log-likelihood ratio (LLR) 
\begin{align}
    \log\frac{Q_{B_i|Y}(0|y)}{Q_{B_i|Y}(1|y)},
\end{align}
which can be calculated from $Q_{X|Y}$ through
\begin{equation}
     Q_{Bi|Y}(b|y)=\sum_{x\in\mathcal{X}^b_i}Q_{X|Y}(x|y),
\end{equation}
where $X_i^b$ is the set of constellation points with the $i$-th label bit equal to $b$, $b \in \{0,1\}$.

According to~\cite{Alvarado15, Bocherer19prob}, an AIR of the system with demapper $Q_{X|Y}$ is
\begin{align}
    \left[\mathrm{H}(X) - \expop[-\log Q_{X|Y}(X|Y)]\right]^+.\label{eq:air}
\end{align}
We now design the communication system to maximize \eqref{eq:air}. As the input entropy $\mathrm{H}(X)$ does not depend on the receiver, the design problem can be rephrased as
\begin{equation}
   \minimize_{\text{Equalizer, Demapper}} \hspace{0.5cm}   \expop [- \log Q_{X|Y}(X|Y)].
   \label{eq:target}
\end{equation}
If the system before the demapper is fixed, then we may minimize over the demapper function $Q_{X|Y}$. If we can optimize both the equalizer and the demapper, then we may express the demapper input as $y=f(r)$, and optimize both over $f$ and $Q_{X|Y}$. In the next section, we discuss several cost functions to optimize either $f$ or $Q_{X|Y}$, or both. Similar techniques have been used to optimize the entire receiver DSP, or even both the transmitter and receiver DSP end-to-end~\cite{Karanov18, Gumus20}.

\section{Cost Functions for Nonlinear Equalizers}
\begin{figure}
    \centering
    \includegraphics[width=0.5\textwidth]{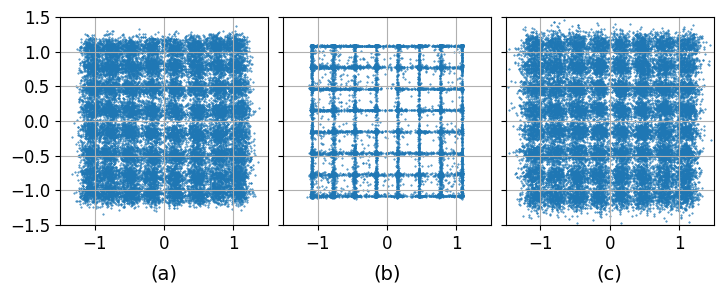}
    \caption{Distibution of the preprocessed received symbol prior to nonlinear equalization (a) and after NN nonlinear equalizer trained with MSE (b) or with regulated MSE (c). }
    \label{fig:scatterplots}
\end{figure}

\textbf{MSE}
The first one~\cite{Jarajreh15, Zhang19, Freire21} minimizes
\begin{equation}
       \text{MSE}(X,f(R)) =  \mathbb{E}[|f(R)-X|^2].
       \label{eq:mse}
\end{equation}
This corresponds to minimizing the squared difference between the equalizer output $Y=f(R)$ and the reference transmit symbols $X$. Minimizing the MSE is not the same as maximizing the AIR, and this can be observed in the equalized constellation in Fig.~\ref{fig:scatterplots}b. The scatterplot after the equalizer is trained is the MSE grid indicating the loss of the soft information.

\textbf{CE} To improve performance, one may realize the equalizer and demapper by a single function. In our model, this corresponds to learning a demapper function $Q_{X|R}$ by minimizing
\begin{equation}
     \text{CE}(X, R) = \expop [- \log Q_{X|R}(X|R)]
\end{equation}
which is the same as \eqref{eq:target} with $R$ replacing $Y$. This approach provides good results, see~\cite{Deligiannidis20, Schadler21}. The drawback is that the trained device acts as an equalizer and soft demapper jointly, i.e., we have no access to an equalized signal for the purposes of carrier and timing recovery.

\textbf{Demapper Proxy} Another solution, proposed in~\cite{Schaedler21soft, bocherer21mlcomm}, optimizes the equalizer based on the demapper output. This is equivalent to solving the following optimization problem:
\begin{equation}
    \minimize_{f, Q_{X|Y}} \hspace{0.5cm}   \expop [- \log Q_{X|Y}(X|f(R))],
    \label{eq:opt_problem}
\end{equation}
where $Y=f(R)$ is the signal after the equalizer function. For practical reasons, the demapper $Q_{X|Y}$ may be parameterized for efficient implementation. For instance, in~\cite{Schaedler21soft}, $Q_{X|Y}$ is parametrized as a max-log approximation (MLA). Note that the optimal $f$ depends on the choice of $Q_{X|Y}$, and the choice of $Q_{X|Y}$ must be taken into account when interpreting the equalized signal $y$.

\section{Entropy-Regularized MSE}
We use the demapper proxy approach and consider the demapper
\begin{align}
Q_{X|Y}(x|y)=\frac{P_X(x)Q_{Y|X}(y|x)}{Q_Y(y)},
\end{align}
where $P_X$ is the input distribution and
\begin{equation}
Q_{Y|X}(y|x)=\frac{1}{2\pi\sigma^2}\exp\left[-\frac{(y-x)^2}{2\sigma^2}\right]
\end{equation}
is a Gaussian channel so that
\begin{equation}
Q_Y(y)=\sum_{x\in\mathcal{X}}P_X(x)Q_{Y|X}(y|x). 
\end{equation}
Note that the demapper is parameterized by the alphabet $\mathcal{X}$ and the noise variance $\sigma^2$. By basic manipulations, the optimization over $f$ becomes
\begin{align}
&\argmin_f \expop [- \log Q_{X|Y}(X|f(R))]\nonumber\\
& =\argmin_f \underbrace{ \expop[|f(R)-X|^2]}_{\text{MSE}(X, f(R))} - \underbrace{ 2 \sigma^2 \expop [ -\log Q_{Y} (f(R))]}_{\text{Entropy regularization}}\nonumber\\
&=: \text{MSE-X}(X, f(R))
\end{align}
This expression is the new proposed cost function for training.
It has two terms: the first is an MSE term as in~\eqref{eq:mse} and the second is an information-dependent term, weighted by a factor proportional to the noise power $\sigma^2$. When the noise power is zero, one recovers the classical MSE. As the noise variance increases, the regularization term preserves the soft information, as can be seen in Fig.~\ref{fig:scatterplots}, where the MSE grid concentrates the equalized signal around the constellation points, corresponding to low entropy, while the regularized MSE maintains a Gaussian-like form.

\section{Experimental Setup}
We test our approach in an experiment.
The channel under test (CUT) carries
an 80GBd dual polarization (DP)-64QAM signal with gross data rate of 960Gb/s. We use 15\% overhead for FEC and 3.47\% overhead for pilots and training sequences, so the net bit rate is 800Gb/s. At the transmitter, a constant amplitude zero auto-correlation (CAZAC) training sequence \cite{pittala2014training} is inserted for frame synchronization, carrier frequency synchronization, and channel estimation.

\tikzset{%
	block/.style = {draw,rounded corners, fill=white, rectangle, minimum height=2em, minimum width=3.5em},
	EDFA/.style = {draw, fill=white, regular polygon, regular polygon sides=3,minimum size=1.1cm},
	sum/.style= {draw, fill=none, circle, node distance=0.5cm,color=black,minimum size=14pt},
	fiber/.style= {draw, fill=none, circle, node distance=2cm,color=red,thick,minimum size=16pt},
}
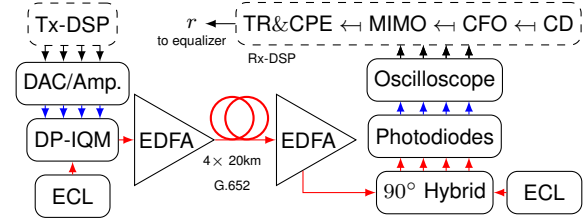
\begin{figure}[h!]
	\centering
	\footnotesize
	\begin{center}
		\begin{tikzpicture}[x=1.5cm, y=1.5cm, >=latex']
			
			\node [block, align = center]  at (0,0) (ECL1) {ECL};
			\node [block,above=0.17cm of ECL1, align = center] (IQM) {DP-IQM};
			\node [block,above=0.17cm of IQM, align = center] (driver) {DAC/Amp.};
			\node [block,dashed,above=0.21cm of driver, align = center] (TXDSP) {Tx-DSP};
			\node [EDFA, right=0.2cm of IQM,align = center,rotate=0,shape border rotate=-90] (EDFA_TX) {\kern-0.5emEDFA\kern-1.0em};
			
			\node [fiber,align = center] at (1.4,0.69) (fiber1) {};	
			\node [fiber, right=-0.4cm of fiber1,align = center] (fiber2) {};
			\node [label, below = 0.1cm of fiber1, draw=none,align = center] (label1) {\tiny 4$ \times$ 20km\\ \tiny G.652 };	
			
			\node [EDFA, right=0.8cm of EDFA_TX,align = center,rotate=0,shape border rotate=-90] (EDFA_RX) {\kern-0.5emEDFA\kern-1.0em};
		
			\node [block,below right=0.15cm and 0.7cm of EDFA_RX, align = center] (HB) {$90^\circ$ Hybrid};
			\node [block,above=0.17cm of HB, align = center] (PD) {Photodiodes};
			\node [block,right=0.17cm of HB, align = center] (ECL2) {ECL};
			\node [block,above=0.17cm of PD, align = center] (Scope) {Oscilloscope};
			\node [block,dashed,above left=0.17cm and -2.83cm of Scope, align = center] (RXDSP) {TR$\&$CPE $\testleft$ MIMO $\testleft$ CFO $\testleft$ CD};
			\node [label, left =0.4cm of RXDSP, draw=none] (label3) {$r$};
			\node [label, below =-0.1cm of label3, draw=none] (label4) {\tiny to equalizer};
			\node [label, below left = 0.0cm and -0.9cm of RXDSP, draw=none] (label1) {\tiny Rx-DSP};
			
			\draw [-latex,color=red] (ECL1) -- (IQM);
			\draw [-latex,color=red] (IQM) -- (EDFA_TX);
			\draw [-latex,color=red] (EDFA_TX) -- (EDFA_RX);
			\draw [-latex,color=red] (EDFA_RX) |- (HB);
			\draw [-latex,color=red] (ECL2) -- (HB);
			
			\draw [-latex] (RXDSP) -- (label3);
			
			\foreach \j in  {0.3,0.1,-0.1,-0.3}
			\draw [-latex, color=black] (Scope.north) + (\j,0) -- ++(\j,0.12);
			
			\foreach \j in  {0.3,0.1,-0.1,-0.3}
			\draw [-latex, color=red] (HB.north) + (\j,0) -- ++(\j,0.12);
			\foreach \j in  {0.3,0.1,-0.1,-0.3}
			\draw [-latex, color=blue] (PD.north) + (\j,0) -- ++(\j,0.12);
			\foreach \j in  {0.3,0.1,-0.1,-0.3}
			\draw [-latex, color=black] (Scope.north) + (\j,0) -- ++(\j,0.12);
			
			\foreach \j in  {0.24,0.08,-0.08,-0.24}
			\draw [-latex, color=black] (TXDSP.south) + (\j,0) -- ++(\j,-0.14);
			
			\foreach \j in  {0.24,0.08,-0.08,-0.24}
			\draw [-latex, color=blue] (driver.south) + (\j,0) -- ++(\j,-0.12);
			
		\end{tikzpicture}
	\end{center}
	\caption{Experimental Setup. Chromatic dispersion (CD) and carrier Frequency offset (CFO) compensation, timing recovery (TR) and carrier phase estimation (CPE).} 
	\label{fig.:Setup}
\end{figure}

Four 120GSa/s digital-to-analog converters (DACs) generate an electrical signal amplified by four 60GHz 3dB-bandwidth amplifiers. A tunable 100kHz external cavity laser (ECL) generates a continuous wave that is modulated by a 32GHz 3dB-bandwidth DP-I/Q modulator. The receiver has an optical $90^\circ$-hybrid and four 100GHz balanced photodiodes. The electrical signals are digitized by an oscilloscope with 256GSa/s and 110GHz 3dB-bandwidth.

\section{Experimental Results}
\newcommand{\nneq}{\text{NN}_\text{eq}}
\newcommand{\nnjointa}{\text{NN}_\text{joint}^1}
\newcommand{\nnjointb}{\text{NN}_\text{joint}^2}
\begin{table}
\centering
\begin{tabular}{lcl}
\toprule
structure & cost function & name\\\midrule
$17|32|26|1$ & MSE, MSE-X & $\text{NN}_\text{eq}$ \\
$17|32|26|3$ &  BCE &$\text{NN}_\text{joint}^1$\\
$17|32|26|\textcolor{red}{16}|3$ & BCE &$\text{NN}_\text{joint}^2$\\\bottomrule
\end{tabular}
\caption{Considered NN structures.}
\vspace{-0.4cm}
\label{tab:nns}
\end{table}
For launch powers $2.7, 6.6, 8.6, 10.7$ dBm, we perform measurements and preprocessing with the setup and Rx DSP displayed in Fig.~\ref{fig.:Setup}. We then train the NNs for every launch power on the first received frame, and we avoid overfitting by evaluating performance on the consecutive frames only. For all equalizers, the input to the first layer is obtained via a symbol-spaced tapped delay line of length $17$. Table~\ref{tab:nns} summarizes the NN parameters.

\textbf{Separate equalization and demapping} The equalizer $\nneq$ has the structure $17|32|26|1$, where each number specifies the number of neurons in the corresponding layer and all hidden layers use ReLU activations. The NN is followed by a demapper with alphabet $\mathcal{X}$ and noise variance $\sigma^2$ estimated from the equalized signal. MSE and MSE-X cost functions are used for training.

\textbf{Joint equalization and demapping} We consider the NNs $\nnjointa$ and $\nnjointb$, see Table~\ref{tab:nns}. The output layers have 3 neurons, one for each bit level. The additional 16-neurons in the hidden layer of structure $\nnjointb$ account for the demapping which requires additional representation capacity. We train the NN using binary CE (BCE).

\begin{figure}
    \includegraphics[width=0.5\textwidth]{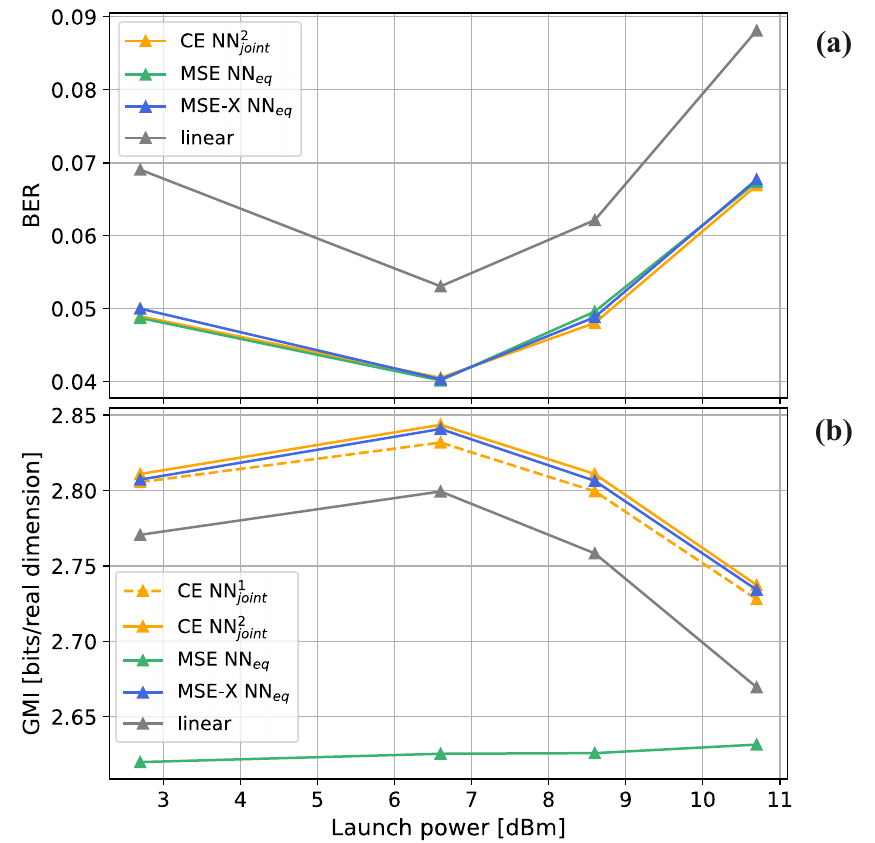}
    \caption{BER (a) and GMI (b) values for increasing launch power.}
    \label{fig:MIplot}
\end{figure}

\textbf{Performance Comparison} 
Fig.~\ref{fig:MIplot}a shows that all NNs achieve the same BER that outperforms the linear equalizer. In particular, $\nneq$ achieves the same BER for MSE and MSE-X.

Fig.~\ref{fig:MIplot}b shows that the three equalizers (1) $\nneq$ trained with MSE-X; (2) $\nnjointa$; and (3) $\nnjointb$ achieve similar GMI. The more complex $\nnjointb$ performance slightly better than $\nneq$, while the less complex $\nnjointa$ performs slightly worse. The AIR degrades when $\nneq$ is trained with MSE, as expected from the MSE grid in the scatterplot.

\section{Conclusions}

Nonlinear equalizers trained with MSE achieve good BER and poor AIR. In this paper, we proposed an entropy-regularized MSE (MSE-X) cost function to train equalizers. MSE-X achieves good BER and good AIR, while still giving access to an equalized signal. Experiments confirmed the practical advantage of MSE-X over MSE.



\newpage
\printbibliography

\vspace{-4mm}

\end{document}